\documentclass[11pt,twoside]{article}

\usepackage{asp2006}
\usepackage{graphicx}
\usepackage{lscape}

\begin{document}
\title{The nature of compact groups of galaxies from
  cosmological simulations}   
\author{Mamon, G. A.}   
\affil{IAP (CNRS \& UPMC), Paris, FRANCE}    
\author{D\'{\i}az-Gim\'enez, E.}
\affil{IATE, Cordoba, ARGENTINA}
\begin{abstract} 
The nature of compact groups (CGs) of galaxies, apparently so dense that the
galaxies often overlap, is still a subject of debate: Are CGs roughly as
dense in 3D as they appear in projection? Or are they caused by chance
alignments of galaxies along the line-of-sight, within larger virialized
groups or even longer filamentary structures?
The nature of CGs is re-appraised using the $z=0$ outputs of three
galaxy formation models, applied to the dissipationless Millennium
Simulation. The same 
selection criteria are applied to mock galaxy catalogs from these models as
have been applied by Hickson and co-workers in redshift space. 
We find 20 times as many mock CGs as the `HCGs' 
found by Hickson within a distance
corresponding to $9000 \, \rm km \, s^{-1}$. This very low (5\%) HCG
completeness is caused by Hickson missing groups that were either faint, near the surface
brightness threshold, of small angular size, or with a dominant brightest
galaxy. 
We find that most velocity-filtered CGs are
physically dense, regardless of
the precise threshold used in 3D group size and line-of-sight elongation, and
of the galaxy formation model used.
This result also holds for mock CGs with the same selection
biases as was found for the HCGs.

\end{abstract}


\section{Introduction: the compact group debate}   

Among the different galaxy environments, compact groups of galaxies
(hereafter CGs) potentially represent 
the densest one involving 4 or more galaxies on small
scales, denser than the cores of rich clusters of galaxies. 
In particular, the catalog of 100 CGs (hereafter, HCGs) 
compiled by \cite{Hickson82} has
generated a strong interest because of its well defined selection criteria:
1) \emph{membership}: at least 4 galaxies within 3 magnitudes from the brightest
one, 2) \emph{compactness}:
mean surface brightness (averaged over the smallest circumscribed circle
containing the galaxy centers, hereafter scc) above a threshold; and 3)
\emph{isolation}: 
no galaxies within 3 magnitudes from the brightest within the
annulus centered on the scc, spanning one to three scc radii. The isolation
criterion effectively rejects clusters of galaxies and makes the HCGs an
environment of its own.

The fundamental question is: are HCGs as dense in 3D as they
appear in projection (e.g. \citealp{HR88}) 
or are they caused by chance alignments of galaxies along the
line-of-sight, within larger groups \citep{Mamon86,WM89} or within longer
filamentary structures \citep{HKW95}?
With mean densities over $10^5$ times the critical density of the
Universe (virialized structures are typically 100 times the critical
density), the crossing times are less than 1 Gyr, and the HCGs ought to be
the ideal environment for galaxy mergers \citep{Mamon92}.
In the  updated HCG catalog with galaxy
redshifts to remove obvious interlopers
\citep{HMdOHP92}, there remained 69 HCGs with at
least four galaxies within $1000 \, \rm km \, s^{-1}$ from the group median.

Still, the identification of CGs in redshift space
does not imply that these structures are dense in real space.
We attempt to respond to this question using state-of-the-art galaxy
formation codes (see details in \citealp{DRMM08}).

\section{Method}    

We build mock CGs in redshift space by 1) starting with the
Millennium dark matter simulation \citep{Springel+05}, 2) building galaxy
catalogs using three different semi-analytical galaxy formation models
(SAMs), by \cite{Bower+06}, \cite{Croton+06}, and \cite{DLB07}, 3)
extracting the mock CGs using the HCG criteria (\citealp{Hickson82}, see \S\,1)
 and  4) applying the velocity filter (see \S\,1) of
\cite{HMdOHP92}. This yields between 3000 and 7000 mock CGs, depending on
  the SAM. We use a bright
galaxy magnitude limit of $R=14.44$, which corresponds to $r_{
\rm SDSS} = 14.77$, so our faint galaxy magnitude limit is $r_{\rm
  SDSS}=17.77$, equal to that of the primary SDSS
spectroscopic sample. 51 HCGs then fully satisfy the HCG selection
criteria with at least 4 velocity concordant galaxies (we rejected 5
non-isolated HCGs noticed by \citealp{Sulentic97}).

\section{Completeness}

\begin{figure}[ht]
\centering
\includegraphics[height=7.5cm]{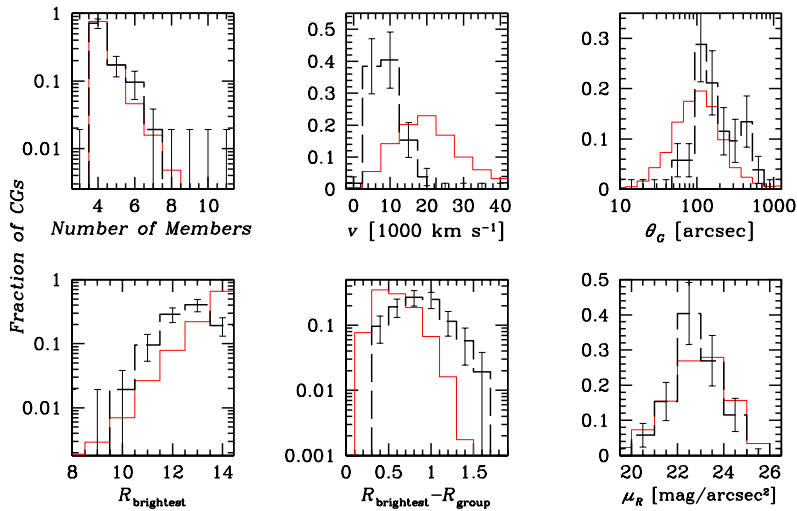}
\caption{Distributions of observables 
in mock De Lucia \& Blaizot (\emph{thin solid histograms}) 
and observed Hickson (\emph{thick dashed histograms}) compact groups,
  both after velocity filtering.
\label{comp1}
}
\end{figure}

In Figure~\ref{comp1}, we show the distribution of observed quantities for
both the observed HCGs and the mock CGs.
One can see that the observed HCGs are progressively more incomplete at
fainter surface magnitudes \citep{WM89}, at larger distances, and for groups
with dominant galaxies (see
\citealp{PIM94}), but also for groups of small angular size.
At a distance of $9000 \, \rm km \, s^{-1}$, 
the mean space density of mock CGs is 20 times that of observed HCGs, hence
the HCG sample at that distance is only 5\% complete (and less so at greater distances).

\section{Nature}

Even with 3-dimensional information, it is not straightforward to decide
which mock CGs are physically dense and which might be caused by chance
projections. For example, one cannot simply separate the mock CGs according
to their binding energy, because the mock CGs are defined with galaxies,
while the intergalactic matter makes up much of the binding energy. Even
if one corrects for this, we have too few galaxies per group for good
measures of binding energy at CG masses below $10^{13.5} M_\odot$.

We use two other statistics to separate dense groups from chance alignments:
the length of the group in 3D (maximum galaxy separation, $s$) and the
line-of-sight (l.o.s.) elongation of the group, defined as $S_\parallel/S_\perp$,
where $S_\parallel$ is the length of the group along the l.o.s., while
$S_\perp$ is the maximum projected separation. Both statistics are computed
on the smallest quartet within the group of 5 or more galaxies (or the group
itself if there are only four members).
\begin{figure}[ht]
\centering
\includegraphics[width=6cm]{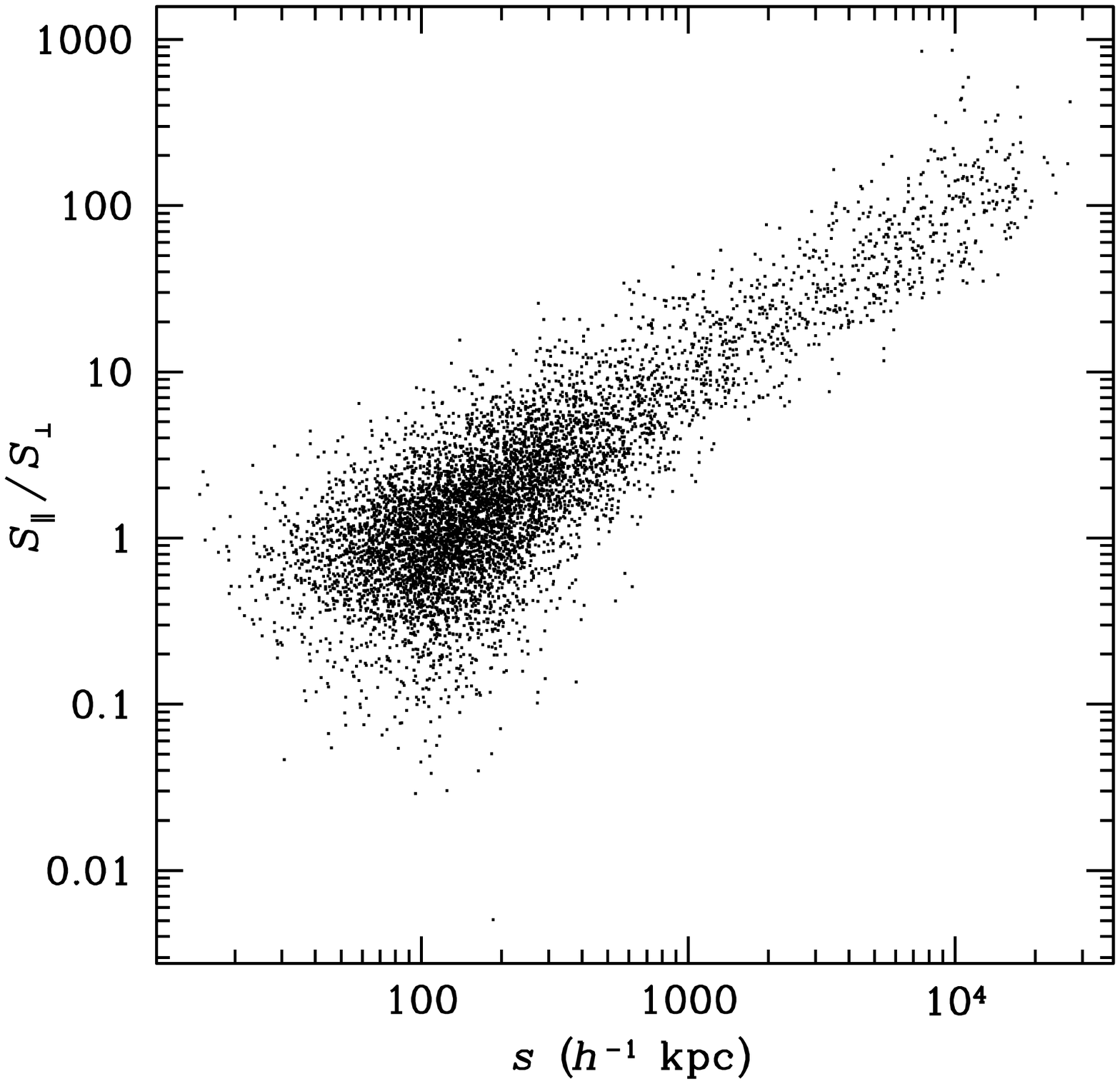}
\includegraphics[width=6cm]{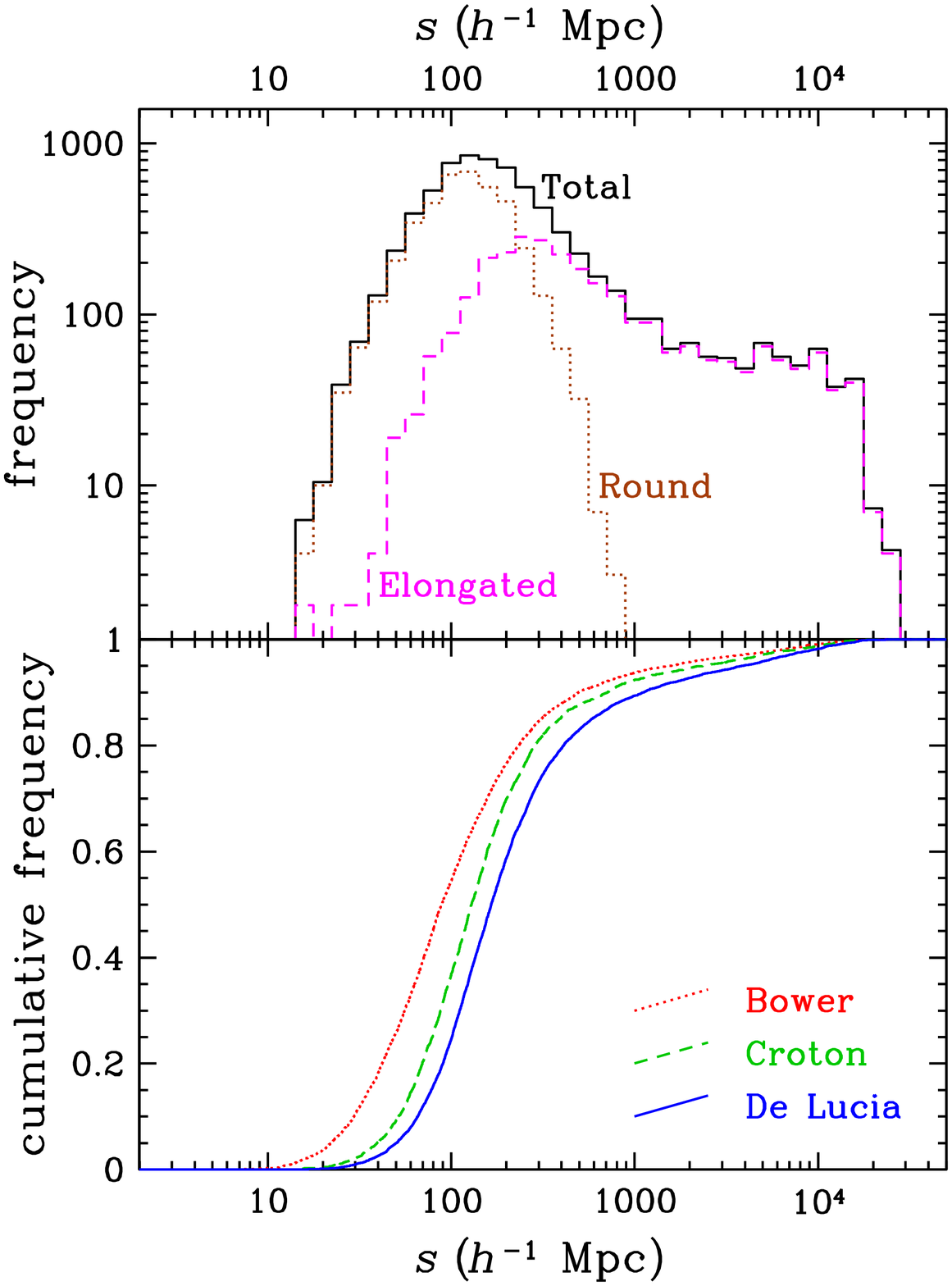}
\caption{\emph{Left}: Line-of-sight elongation vs. 3D length for smallest
  quartets within 
mock (De Lucia \& Blaizot [DLB]) velocity-filtered (v.f.) CGs.
\emph{Right}:
Differential (\emph{top} for DLB) and cumulative (\emph{bottom})
  distributions of 3D length of v.f.
CGs. The \emph{dotted} and \emph{dashed histograms}
  show the round ($S_\parallel/S_\perp<2$) and elongated
  ($S_\parallel/S_\perp>2$) subsamples.
\label{sdis}
}
\end{figure}
Figure~\ref{sdis} shows these measures of group size and l.o.s. elongation.
Still, it is not clear where to make cuts in, say, $s$ vs
$S_\parallel/S_\perp$.
We consider conservative criteria for physically dense groups:
short ($S<200 \, h^{-1} \, \rm kpc$),
round ($S_\parallel/S_\perp<2$),
or combinations of both.
%
%
\begin{table}
\begin{center}
\caption{Fraction of velocity filtered mock CGs that are physically dense
\label{nattab}}
\begin{tabular}{lccc}
\hline
 & \multicolumn{3}{c}{galaxy formation model} \\
\cline{2-4}
Criterion & Bower+06 & Croton+06 & De Lucia+07 \\
\hline
$s < 200 \, h^{-1} \, \rm kpc$ & 0.77 & 0.70 & 0.59 \\
$S_\parallel/S_\perp < 2$ & 0.70 & 0.67 & 0.59 \\
$s < 100 \, h^{-1} \, \rm kpc \hbox { OR}$ \\
$(s < 200 \, h^{-1} \, \rm kpc \hbox { AND } S_\parallel/S_\perp < 2)$ & 0.71
& 0.63 & 0.52 \\
\hline
\end{tabular}
\end{center}
\end{table}
Table~\ref{nattab} indicates that regardless the criterion and the SAM used,
most of the mock CGs are physically dense.
Similar fractions are returned using samples of mock CGs built with the same
selection biases as the HCGs.

\section{Discussion}

A similar work has been performed by \cite{McCEP08}, who defined physically
dense groups as those that are found with a Friends-of-Friends linking length
of $200 \, h^{-1} \, \rm kpc$. \citeauthor{McCEP08} find that 35\% 
of their mock compact groups (before velocity filtering) are physically
dense, while we typically find 20\% with the \citeauthor{DLB07} model used by them.

\acknowledgements
We acknowledge the Millennium Simulation, and public galaxy formation outputs
by Bower et al., Croton et al. and De Lucia \& Blaizot, on top of it. 
G.A.M. thanks the Dept. of Physics at the Univ. of Oxford for
hospitality while part of this research was performed.



\end{document}